\documentclass[10pt]{revtex4}
\usepackage{graphicx}
\usepackage{amsmath,amssymb}
\usepackage{mathtools}

\begin{document}

\title{The Quantum Double Slit Experiment With Local Elements of Reality}
\author{Vlatko Vedral}
\affiliation{Clarendon Laboratory, University of Oxford, Parks Road, Oxford OX1 3PU, United Kingdom\\Centre for Quantum Technologies, National University of Singapore, 3 Science Drive 2, Singapore 117543\\
Department of Physics, National University of Singapore, 2 Science Drive 3, Singapore 117542}

\begin{abstract}
We present a fully local treatment of the double slit experiment in the formalism of quantum field theory. Our exposition is predominantly pedagogical in nature and exemplifies the fact that there is an entirely local description of the quantum double slit interference that does not suffer from any supposed paradoxes usually related to the wave-particle duality. The wave-particle duality indeed vanishes in favour of the field picture in which particles should not be regarded as the primary elements of reality and only represent excitations of some specific field configurations. Our treatment is general and can be applied to any other phenomenon involving quantum interference of any bosonic or fermionic field, both spatially and temporally. For completeness, we present the full treatment of single qubit interference in the same spirit. 
\end{abstract}

\maketitle

Feynman observed that the two-slit interference experiment "...contains the only mystery" of quantum mechanics \cite{Feynman}. This is because any, more complicated, interference phenomenon can always be built up from the interferences of two level systems. We therefore start our discussion with the usual quantum optical version of the Young double slit experiment and discuss various generalisations later. 

Light is an electromagnetic wave. As such it obeys Maxwell's equations which lead to a wave equation for both the electric and the magnetic field components (classically, the electromagnetic field consists of six numbers assigned to every point in space and at every instance of time). The quantum ``twist" is to upgrade the electric and the magnetic field components (from mere numbers) into six (pairwise and in a specific fashion) non-commuting operators \cite{Weinberg}. The key, however, is that these operators still obey the same aforementioned equations. Therefore, in order to study any quantum interference phenomenon, we can utilise all the mathematics pertaining to the classical waves, such as the Huygens principle and the Kirchhoff diffraction formula, and only remember that the quantities evolving in this manner are actually quantum operators or q-numbers (and not the c-numbers representing the classical electromagnetic field). From this, it is already clear that the natural picture to work in will be the Heisenberg representation of quantum dynamics in which the relevant operators evolve in time while the state always remains the same. Suffice it to say, any other picture of quantum dynamics will do just as well, though other representations could be considered less ``natural" based on the presented logic. 

With this in mind, we proceed to discuss the double slit experiment involving a single photon impinged on a grating and otherwise in the vacuum. We deliberately focus on the single photon as this is the ``most quantum" that the electromagnetic field can get. However, given that we will be speaking the language of Heisenberg, most of our treatment will automatically apply to any other state of the field. It is just that when it comes to calculating the probability for the photon to end up at a point in the far field (we will assume the Fraunhoffer limit without any loss of generality), we will ultimately perform the averaging with respect to the single photon state. 

Since the photon polarization is irrelevant, we can treat the electromagnetic field as being a scalar and, instead of tracking the six operators, it will be sufficient to use only one, $\Psi (x,t)$. The whole point of the calculation is to relate the number operator at the source, $S$, (at the time when a single photon is emitted), to the number operator at the point of detection, $D$, when this photon is detected in the far field. This addresses the question: how is the local element of reality representing the intensity of light at the source related to the element of relativity representing the intensity of light at the detector? The number operator in question is $\Psi^{\dagger} (x,t)\Psi (x,t)$. As is customary in classical wave optics, we specialise in a single frequency of light, which leads us from the wave equation to the Helmholtz equation (thereby factoring out the temporal part of the phase $e^{-i\omega t}$). It is well know that the Helmholtz equation can be solved using the methods of Kirchhoff \cite{Born}. 

Suppose for simplicity that the source is equidistant from the two (point-like) slits. Then the quantum field at the detection point can be related to the quantum field at the source like so:
\begin{equation}
\Psi (x_D) = \left(\frac{e^{ikr_1}}{r_1} + \frac{e^{ikr_2}}{r_2}\right)\frac{e^{iks_0}}{s_0}\Psi (x_S)
\end{equation}
where $s_0$ is the distance from the source to the slits and $r_1$ and $r_2$ are the distances from the slits to $D$. This is the same as the standard classical wave optics formula for interference fringes, other than the fact that $\Psi$ is a quantum field operator (proportional to the sum of the creation and annihilation operators, $a + a^{\dagger}$). In order to compute the interference fringes we need to relate
the number operator at $D$ to the number operator at $S$. Taking the dagger of the above formula and multiplying the two together we obtain
\begin{equation}
\Psi^{\dagger} (x_D) \Psi (x_D) = \left|\left(\frac{e^{ikr_1}}{r_1} + \frac{e^{ikr_2}}{r_2}\right)\right|^2\Psi^{\dagger} (x_S)\Psi (x_S)
\end{equation}
where we have omitted the $1/s^2_0$ factor since it is just part of the overall normalisation. The restriction to the state with one photon now instructs us to reduce this expression to 
\begin{equation}
|1_D\rangle\langle 1_D| = \frac{(1+\cos (k(r_1-r_2))}{2} |1_S\rangle\langle 1_S|
\end{equation}
where we have assumed that $r_1\approx r_2$ in the denominator, which we have then omitted as part of the overall normalisation. 
This is the first time in our calculation that we have to invoke quantum states (and we do so thinking of them as the eigenstates of the corresponding operators, so that $\Psi^\dagger \Psi = \sum_{n=0}^{\infty} n |n\rangle\langle n|$).
Taking the expected value with respect to $|1_S\rangle$ leads us to the (not properly normalised) probability to detect a photon at $D$
\begin{equation}
|\langle 1_D|1_D\rangle|^2 = \frac{(1+\cos (k(r_1-r_2))}{2} 
\end{equation}
which is the standard double slit pattern of fringes in the far-field limit and assuming that the slits are point-like.  

Several comments are now in order. First, the quantum field description inherits all the local properties of classical fields. By that we mean that the elements of reality, in this case the operators pertaining to the electromagnetic field and defined at each point \cite{Vedral-local}, cannot affect one another at a distance and instantaneously. There is no action at a distance. Namely, changing one operator at one point in space by actions local to that point, leaves all other operators intact \cite{Vedral-mach}. Micro-causality is also satisfied (it is a weaker form of locality than the previous one), meaning that operators at space-like separated points also commute, or that any signal cannot propagate faster than the speed of light. This way, quantum field theory complies with relativity. 

Second, this field-theoretic picture must give up the idea that particles are elements of reality, even in the case of the experiment involving only one photon! One way of seeing that is to write the intermediate state of the photon at the two slits (labeled as $1$ and $2$) as an entangled state of the form $|0_1,1_2\rangle + |1_1,0_2\rangle$ (in the Schr\"odinger picture). For us to talk about the photon as existing either at slit one or at slit two (but not in a superposition of both) would consitute a local hidden variable model, which we know not to be consistent with quantum physics \cite{Vedral,Dunningham}. This resolves the wave-particle duality since the objects that interfere are the quantum fields, not particles. 

Finally, it is clear that even though we were describing the diffraction of the quantized electromagnetic field, our description applies to any other type of a particle. We could have imagined a single electron diffraction at a crystal and the calculation would virtually be the same. The only difference is that the corresponding operators would anti-commute, which does not affect the single particle interference. Every particle in quantum field theory is an excitation of its own field and is therefore mathematically treated in an analogous way. And, needless to say all other states of the quantized field operate the same way. A similar view, that the fields are the ultimate entities and not the particles, has been expressed by Hobson \cite{Hobson}. 

Since all complex quantum phenomena are made up of quantum bits we now conclude by presenting the general treatment of the single qubit interference in the spirit of the above discussion. Any more complicated interference experiments can be constructed out of qubits (be they related to spatial or temporal interference). In the above example, the states involving no photons and one photon could be thought of as two orthogonal qubit states and so can the two paths through the slits. 

The key quantity to compute is the probability amplitude to go from one eigenstate of one qubit operator and at one time to another eigenstate of another qubit operator at a later time.  Let us assume (without loss of generality) that the Hamiltonian is given by $H = \frac{\hbar\omega}{2} \sigma_Z$.  We can always express this with the ``electric"-like and ``magnetic"-like observables by associating two quantum harmonic oscillators to the qubit (the procedure sometimes refered to as the second quantisation):
\begin{equation}
a^\dagger \sigma_Z a = a^{\dagger}_x a_x - a^{\dagger}_y a_y
\end{equation}
where $a_{x} = (x+ip_x)/\sqrt{2}$ and $a_{y} = (y+ip_y)/\sqrt{2}$ and $a=(a_x,a_y)$. This is tantamount to second-quantizing the qubit and acknowledging that each of the two levels could be occupied by an arbitrary number of particles (such that the particles belonging to the $x$ oscillator occupy the excited state while the particles pertaining to the $y$ oscillator occupy the ground state). The other two Pauli operators are mapped as $a^\dagger\sigma_X a= a^\dagger_{x}a_{y}+a^\dagger_{y}a_{x}$ and $a^\dagger\sigma_Y a= i (a^\dagger_{x}a_{y}-a^\dagger_{y}a_{x})$. 

We follow the road that leads most naturally to the Heisenberg picture of quantum dynamics. In that case the Hamiltonian can be written as: 
\begin{equation} 
H = \frac{\hbar\omega}{2} (a^{\dagger}_x a_x - a^{\dagger}_y a_y) = \frac{p^2_x}{2}+\frac{\omega^2 x^2}{2} - \frac{p^2_y}{2} - \frac{\omega^2 y^2}{2}
\end{equation}
The $x,y,p_x,p_y$ operators evolve according to the Hamilton's equations of motion:
\begin{equation} 
\frac{\partial x}{\partial t} = \frac{\partial H}{\partial p_x} \;\;\;\;\; \frac{\partial p}{\partial t} =- \frac{\partial H}{\partial x}
\end{equation}
and the same for the $y$ oscillator. The derivative of the $H$ operator with respect to the $p$ operator is defined as 
\begin{equation} 
\frac{\partial H}{\partial p_x} = \lim_{\epsilon \rightarrow 0} \frac{H(...,p+\epsilon I,...)-H(...,p,...)}{\epsilon}
\end{equation}
We are now in the position to write down the (Heisenberg) equations of motion for all the observables. We obtain:
\begin{equation}
\dot x = \frac{\hbar\omega}{2} p_x \;\;\;\; \dot p_x = - \frac{\hbar\omega}{2} x
\end{equation}
and similarly for the $y$ oscillator. These can easily be solved. We can then work backwards to obtain the equations of motion for the Pauli qubit operators:
\begin{equation} 
\sigma_X (t) = \cos (\omega t) \sigma_X (0) + \sin (\omega t) \sigma_Y (0) 
\end{equation}
and similarly for $\sigma_Y$ (both first and second quantized). The operator $\sigma_Z$ is a constant of motion, i.e. $\sigma_Z (t)= \sigma_Z (0)$ (as seen from the fact that $\sigma_Z$ commutes with the Hamiltonian).
If $\sigma_X(0)=\sigma_X$ we can then inquire, for instance, about the amplitude to go from the say $|+\rangle$ eigenstate at time zero to the $|-\rangle$ eigenstate at time $t$. It is clear that the $X$ component of the qubit oscillates periodically between $\sigma_X$ and $\sigma_Y$ and that this allows us to calculate all experimentally accessible quantities. 

Here, of course, the (second quantized) qubit is exhibiting interference in time, unlike the photon in the double slit experiment which could be said to interfere in ``space". Of course, the terminology here is somewhat arbitrary (since the spatial and the temporal interference are not relativistic invariants) but 
in general all perturbations, such as spatial translations, rotations and temporal translations, can be included \cite{Schwinger} and treated on an equal footing using the quantum action principle. This is done by saying that the variation of the amplitude between the initial (at $t_1$) and the final state (at $t_2$) is written as
\begin{equation}
\delta \langle b,t_2|a,t_1\rangle = \langle b,t_2|(p\delta x - H\delta t)|^2_1 |a,t_1\rangle
\end{equation}
and this identity takes into account both the spatial as well as the temporal variation.  The two paths in the double slit experiment can be represented by the two eigenstates of $\sigma_Z$, say, and its second quantized form simply implies that we can have any number of particles in each. And of course, the phase $kr$ that govern spatial interference now becomes the temporal component $\omega t$. 

Finally, the claim that the double slit experiment demonstrates quantum non-locality, seems to stem from the usage of the first quantized treatment \cite{Aharonov} and the fact that in such a treatment there are no fields to mediate causal propagation of effects between different spacetime points. The same applies to more complicated phenomena such as the Aharovon-Bohm effect \cite{ABMAVE}. Quantum field theory is not just a fantastically accurate calculational tool. It may well be plagued by all sorts of troublesome divergences, however, its successes at resolving some key foundational issues at the heart of quantum mechanics are well worth remembering from time to time.

\textit{Acknowledgments}: The author thanks Chiara Marletto for useful comments. He also acknowledges funding from the National Research Foundation (Singapore), the Ministry of Education (Singapore) and Wolfson College, University of Oxford.

\end{document}